\documentclass{nature}
\usepackage{psfig}

\title{The magnetic nature of disk accretion onto black holes}

\author{Jon M. Miller$^1$,$~~$John Raymond$^{2}$,$~~$Andy Fabian$^3$,
  $~~$Danny Steeghs$^2$,$~~$Jeroen Homan$^4$,\\
  $~~$Chris Reynolds$^5$,$~~$Michiel van der Klis$^6$,$~~$ Rudy
  Wijnands$^6$}

\long\def\symbolfootnote[#1]#2{\begingroup%
\def\thefootnote{\fnsymbol{footnote}}\footnote[#1]{#2}\endgroup} 
\def\blfootnote{\xdef\@thefnmark{}\@footnotetext}
 
\def\gtsima{$\; \buildrel > \over \sim \;$}
\def\ltsima{$\; \buildrel < \over \sim \;$}
\def\gsim{\lower.5ex\hbox{\gtsima}}
\def\lsim{\lower.5ex\hbox{\ltsima}}
\def\simgt{\lower.5ex\hbox{\gtsima}}
\def\simlt{\lower.5ex\hbox{\ltsima}}
\def\simpr{\lower.5ex\hbox{\prosima}}

\begin{document}

\maketitle
\small
\begin{affiliations}

\item University of Michigan Department of Astronomy, 500 Church
Street, Ann Arbor, MI 48109, USA
\item Harvard-Smithsonian Center for Astrophysics, 60 Garden Street,
Cambridge, MA, 02138 USA
\item University of Cambridge Institute of Astronomy, Cambridge CB3 OHA, UK
\item Kavli Institute for Astrophysics and Space Research,
Massachusetts Institute of Technology, 77 Massachusetts Avenue,
Cambridge, MA 02139, USA
\item University of Maryland Department of Astronomy, College Park, MD
20742, USA
\item Astronomical Inst. Anton Pannekoek, Univ. of
Amsterdam, Kruislaan 403, 1098 SJ Amsterdam, NL
\end{affiliations}
\normalsize

\begin{abstract}
Although disk accretion onto compact objects - white dwarfs, neutron
stars, and black holes - is central to much of high energy
astrophysics, the mechanisms which enable this process have remained
observationally elusive.  Accretion disks must transfer angular
momentum for matter to travel radially inward onto the compact
object\cite{ss}.  Internal viscosity from magnetic
processes\cite{ss}$^,$\cite{bh91}$^,$\cite{hgb95}$^,$\cite{bh98} and
disk winds\cite{bp} can in principle both transfer angular momentum,
but hitherto we lacked evidence that either occurs.  Here we report
that an X-ray-absorbing wind discovered in an observation of the
stellar-mass black hole binary GRO J1655$-$40\cite{orosz} must be
powered by a magnetic process that can also drive accretion through
the disk.  Detailed spectral analysis and modeling of the wind shows
that it can only be powered by pressure generated by magnetic
viscosity internal to the disk or magnetocentrifugal forces.  This
result demonstrates that disk accretion onto black holes is a
fundamentally magnetic process.
\end{abstract}
\normalsize

To study the nature of disk accretion onto black holes, we observed
the transient source GRO~J1655-40 with the Chandra X-ray Observatory
for 63.5 ksec starting on 1 April 2005 at 12:41:44 (TT), during an
X-ray bright phase of its 2005 outburst.  GRO~J1655$-$40 is a binary
system at a distance of 3.2~kpc that harbors a black hole with a mass
of $7.0~{\rm M}_{\odot}$, which accretes from an F3 IV -- F6 IV star
with a mass of $2.3~{\rm M}_{\odot}$ in a 2.6-day orbit\cite{orosz}.
The inner disk is viewed at an inclination of 67--85$^{\circ}$ (nearly
edge-on)\cite{orosz}$^,$\cite{hr}.  Using Chandra, we obtained a
robust high resolution X-ray spectrum of GRO~J1655$-$40 in the
soft X-ray band (see Figure 1).

It is common to decompose the broad-band spectra of stellar-mass black
holes into disk blackbody and power-law components.  Assuming the
standard equivalent neutral hydrogen absorption\cite{dl} along the
line of sight to GRO~J1655$-$40 ($N_{H} = 7.4\times 10^{21}~{\rm
atoms}~{\rm cm}^{-2}$), this spectral model gives a disk temperature
of $kT = 1.34(1)$~keV and a photon power-law index of $\Gamma =
3.54(1)$ in the 1.2--19~\AA~ range, and a total flux of $4.70(5)\times
10^{-8}~{\rm erg}~{\rm cm}^{-2}~{\rm s}^{-1}$ (unabsorbed).  This flux
implies a luminosity of $L = 3.3\times 10^{37}~{\rm erg}~{\rm s}^{-1}$
for $d=3.2$~kpc, or 4\% of the Eddington limit for a $7~{\rm
M}_{\odot}$ black hole.  The disk contributes 65\% of the unabsorbed
flux.

The HETGS spectra of GRO~J1655$-$40 contain 90 absorption lines
significant at the 5$\sigma$ level of confidence or higher.  We can
confidently identify 76 of these lines with resonance lines expected
from over 32 charge states.  The properties of these lines were
measured with simple Gaussian line functions and local continuum
models.  Line centroid wavelengths and oscillator strengths were taken
from a set of standard
references\cite{vvf}$^,$\cite{nist}$^,$\cite{nah}.  Two findings
demand that the absorption arises in a disk--driven wind.  First, the
lines show blue-shifts in the 300--1600~${\rm km}~{\rm s}^{-1}$ range,
indicating a flow into our line of sight (see Figure 1, Figure 2, and
the Supplementary Information).  Second, the spectra contain no strong
emission lines; this fact signals that the absorbing gas is mostly
equatorial along the plane of the disk.

To better understand the nature of the wind, we constructed a
number of photoionized plasma models based on the methodology and
atomic physics packages described in a prior
work\cite{jr93}$^,$\cite{m04}.  (The models used in this work differ
only in that new dielectronic recombination rates for Fe and Ni were
used to more accurately describe L-shell ions\cite{colgan}.)  These
models describe a gas in photoionization equilibrium, with heating by
photoionization and Compton scattering and cooling by line and
continuum emission and Compton scattering.  The predicted ionization
state of the gas is combined with a curve of growth\cite{sp78} for a
chosen velocity width to determine line equivalent widths using Voigt
profiles.  The code accounts for line blends self-consistently.  A
standard set of elemental abundances are used in the code\cite{grev}.
The illuminating spectrum is taken to be the composite thermal and
non-thermal continuum spectrum described briefly above.

The observed spectrum is consistent with absorption in a
constant-density slab with a thickness of approximately $2.5 \times
10^{8}~{\rm cm}$ and a number density of $n = 5.6 \times 10^{15}~{\rm
atoms}~{\rm cm}^{-3}$ at a mean distance of $4.8 \times 10^{8}~{\rm
cm}$ from the black hole.  This translates to approximately
$200~R_{Schw.}$ (where $R_{Schw.} = 2GM/c^{2}$); the component of the
wind velocity in our line of sight is much slower than the local
virial velocity.  The temperature of the gas is 0.2--1.0$\times
10^{6}$~K.  An intrinsic FWHM of $300~{\rm km}~{\rm s}^{-1}$ matches
the observed lines which are not saturated (see Figure 2 and the
Supplementary Information).  An important feature of this wind is that
it is highly ionized: ${\rm log}(\xi) = {\rm log}(L_{X}/nr^{2})
=$~4.2--4.7 (where $\xi$ is the ionization parameter, $L_{X}$ is X-ray
luminosity, $n$ is density, and $r$ is the radius from the ionizing
source).

The data permit strong independent constraints on the geometry and
extent of the wind absorption.  An inner radial extent of
$10^{7.5}$~cm is set by the density at which Fe~XXIII lines are
produced.  At smaller radii, the ionization parameter requires a
density so large that the metastable 2s2p$^3{\rm P}$ level of Fe~XXIII
would be populated, producing lines that are not observed.  An outer
radial extent of $10^{9.5}$~cm is set by dilution and line ratios.  At
larger radii, the thickness of the absorbing gas becomes large
compared to the distance, so $1/r^{2}$ dilution of the radiation field
makes it impossible to get enough total column at high enough
ionization parameter.  A lower limit of 6$^{\circ}$ on the height of
the gas above the disk midplane comes from the fact the line of sight
must pass over the outer edge of the disk\cite{vrt}.  An upper limit
on the vertical extent comes from the lack of emission features.  A
Monte Carlo simulation for a cascade resulting from absorption in the
2s-4p line of Fe XXIV predicts a 2p-3s emission line with equivalent
width $0.65\times 28\Omega / 4\pi = 18 \Omega/ 4\pi$~m\AA~ (where
$\Omega$ is the solid angle covering factor).  An upper limit of a
2m\AA~ to such a feature implies an upper limit of about 12$^{\circ}$
to the vertical extent of the absorbing gas.

The properties of thermally--driven winds have been studied,
especially within the context of outer accretion disks being
illuminated by a central X-ray source\cite{beg83}.  In such cases, it
is possible to define a critical radius $R_{C} = (1.0\times
10^{10})\times (M_{BH}/M_{\odot})/(T_{C8})~{\rm cm}$, and a wind may
occur for any $R/R_{C} > 0.1$ (where $T_{C8}$ is the gas temperature
in units of $10^{8}$~K).  Based on temperatures derived from our
photoionization models, the smallest possible value of $R_{C}$ for the
wind observed in GRO~J1655$-$40 is $7 \times 10^{12}~{\rm cm}$.  Thus,
the minimum radius at which a disk wind can be thermally driven is
approximately two orders of magnitude greater than is plausible in
GRO~J1655$-$40.  Similar results are obtained when our results are
compared to new models for the winds in AGN such as NGC~3783\cite{cn05}.

The wind observed is too highly ionized to be driven by radiation
pressure.  The overwhelming majority of absorption lines observed are
from He-like and H-like species, which means that there is little UV
opacity in the wind by which momentum may be transferred.  At
ionization parameters of $\xi > 10^{3}$, models for line--driven winds
in AGN indicate that UV emission lines provide no additional driving
force\cite{psk00}.  Moreover, our photoionization code
measures radiation pressure as the momentum of the photons absorbed,
which is comparable to the electron scattering radiation pressure, and
even together these effects fall far short of producing the observed
momentum flux in the wind.

Given that thermal and radiative driving fail by orders of magnitude,
magnetic driving of the wind in GRO~J1655$-$40 is the only plausible
mechanism remaining.  Our model implies a mass loss rate in the wind
of $\dot{m}_{w} = 3.5\times 10^{17}$~g/s or approximately 0.5~g/${\rm
cm}^{2}$/s.  For a typical blue-shift of 500~km/s, this translates
into a kinetic energy flux of $6.3\times 10^{14}$~erg/${\rm
cm}^{2}$/s.  An angle of $9^{\circ}$ above the disk midplane at radius
of $4.8\times 10^{8}$~cm from the black hole corresponds to a height
of $Z=0.15r$; an energy flux of $2.0\times 10^{16}$~erg/${\rm
cm}^{2}$/s is required to lift the gas to that height.  The luminosity
of the central engine as derived from fits to the continuum implies a
mass accretion rate of $\dot{m}_{a} = 3.7\times 10^{17}$~g/s for an
accretion efficiency of 10\%.  The viscous energy flux dissipated is
given by $3GM_{bh} \dot{m}_{a} / 4\pi r^{3} = 2.3 \times
10^{18}$~erg/${\rm cm}^{2}$/s.  Recent simulations show that the
magneto-rotational
instability\cite{bh91}$^,$\cite{hgb95}$^,$\cite{bh98} can not only
drive turbulence, viscosity, and accretion through a disk, but can
transmit 25\% of the magnetic energy flux vertically out of the
disk\cite{ms00} and drive a wind.  In the case of GRO~J1655$-$40, 25\%
of the viscous energy flux is comparable to the flux required to drive
the wind to infinity.  The wind outflow velocity and mass outflow rate
are remarkably similar to predictions resulting from from simulations
of magnetically--driven winds from MRI disks\cite{proga03}.

It is also possible that the wind is driven by magnetocentrifugal
forces\cite{bp}.  The absorption spectrum contains no information
about velocities tangential to our line of sight.  Theoretical models
show that largely equatorial winds can arise via magnetocentrifugal
driving when the magnetic field vector makes a small angle to the disk
plane\cite{spruit}; however, recent simulations suggest that it is
difficult to maintain the poloidal magnetic field required in a wind
with a large mass outflow\cite{proga03}.  Given that the wind arises
in a disk which is almost certainly Keplerian, however,
magnetocentrifugal driving cannot be discounted.  The winds in some
young stars (FU Orionis class) are driven by magnetocentrifugal
forces\cite{calvet}.  Tapping into the rotational velocity of the disk
would help to expel the wind to infinity.  Although internal magnetic
viscosity and magnetic winds are sometimes posited as complete and
separate processes, the most physically realistic scenario may be one
in which both processes act to drive disk accretion and outflows.

Winds are commonly observed in accreting compact objects; however,
GRO~J1655$-$40 is the first case wherein it is clear that the wind
must be launched from the disk (not the companion star) and must be
driven primarily by magnetic processes (not thermal and radiative
pressure).  An X-ray wind with some similarities to those in AGN was
detected in the accreting neutron star binary Circinus X-1\cite{sb02};
however, in Circinus X-1 contributions from a massive companion star
cannot be ruled out, and the wind can be driven by thermal and
radiation pressure\cite{sb02}.  A wind more certainly tied to the disk
was detected in the black hole binary GX~339$-$4, but too few lines
were detected to constrain the driving mechanism\cite{m04}.  Though
radiative driving may be important in most white dwarf systems, there
is at least one system where it may be inadequate\cite{mauche}.
Similarly, theoretical considerations suggest that magnetocentrifugal
forces are important in AGN winds\cite{kk94}$^,$\cite{e05}, but
present data do not yet require or rule out these effects.  Our
results therefore represent a rare and crucial insight into the nature
of the processes which drive disk accretion in black holes.  Magnetic
pressure supplied by the disk provides a natural means of driving the
highly ionized winds observed in many accreting stellar-mas black
holes and neutron stars, and may play a role in driving the hottest
winds observed in AGN.  Indeed, winds and jets are ubiquitous features
in accretion-powered astrophysics, and the role of magnetic processes
revealed in GRO~J1655$-$40 gives a broad observational insight into
the physical coupling between inflows and outflows in accreting
compact objects.

\noindent
Correspondence and requests for materials should be addressed to
J.M.M. (jonmm@umich.edu)  

\noindent
We acknowledge insightful conversations with Nuria Calvet, Lee
Hartmann, Daniel Proga, and Michael Rupen.  We are indebted to Andrea
Prestwich, Harvey Tananbaum, and the Chandra staff for its help in
making this observation possible.  We thank Leslie Sage and Beth
Lauritsen for editorial insights. This work was supported by funding
from NASA through the Chandra guest obsever program (to J. M. M.).

\newpage



Figure 1.  The slice of the disk wind spectrum observed in
GRO~J1655$-$40 with Chandra.  In the plot above, the best-fit
phenomenological model is shown in blue.  The model in red plots the
natural line wavelengths, to illustrate that the observed absorption
is blue-shifted.  The errors shown are 1$\sigma$ statistical errors on
the photon flux.  To obtain this high resolution spectrum, the Chandra
High Energy Transmission Grating Spectrometer (HETGS) was used to
disperse the X-ray flux onto the Advanced CCD Imaging Spectrometer
(ACIS), which was operated in ``continuous-clocking'' mode.  The data
reduction and preparation was performed using the latest version of
the standard Chandra packages (``CIAO''), and a procedure typical for
this instrumental configuration\cite{m04}.  The spectra presented in
this work were produced by adding the first-order spectra from the
HETGS medium energy grating (MEG) at a resolution of 0.005~\AA~ per
bin, and the first-order spectra from the high energy grating (HEG) at
a resolution of 0.0025~\AA~ per bin.  The HEG spectrum was used to
characterize the 1--11\AA~ band, and the MEG spectrum was used to
characterize the 11--19\AA~ band.  The continuum emission was
characterized using the HEG spectrum.  All spectral fits were made
using the ISIS\cite{houck} spectral fitting package.

Figure 2.  Comparison of the best model for the disk wind in
GRO~J1655$-$40 to the data.  The plot above shows the ratio of the
absorption line equivalent widths measured in GRO~J1655$-$40, to the
equivalent widths predicted by the photoizonization model which best
describes the disk wind.  The model assumes an internal velocity
widths of 300~km/s.  Revised solar abundances\cite{grev} are adequate
to describe the elements between Na and K (inclusive).  Abundances of
twice the revised solar value are required to describe the lines
observed rom other elements.  The full spectrum of GRO~J1655$-$40 and
the measured properties of the absorption lines are detailed in the
Supplemental Information.




\clearpage


\begin{figure*}
\vspace{-2cm}
\hspace{-1cm}
\centerline{\psfig{figure=2005-12-14717_fig1.ps,width=12.0cm}}
\end{figure*}
\begin{figure*}
\end{figure*}
\clearpage


\begin{figure*}
\vspace{-2cm}
\hspace{-1cm}
\centerline{\psfig{figure=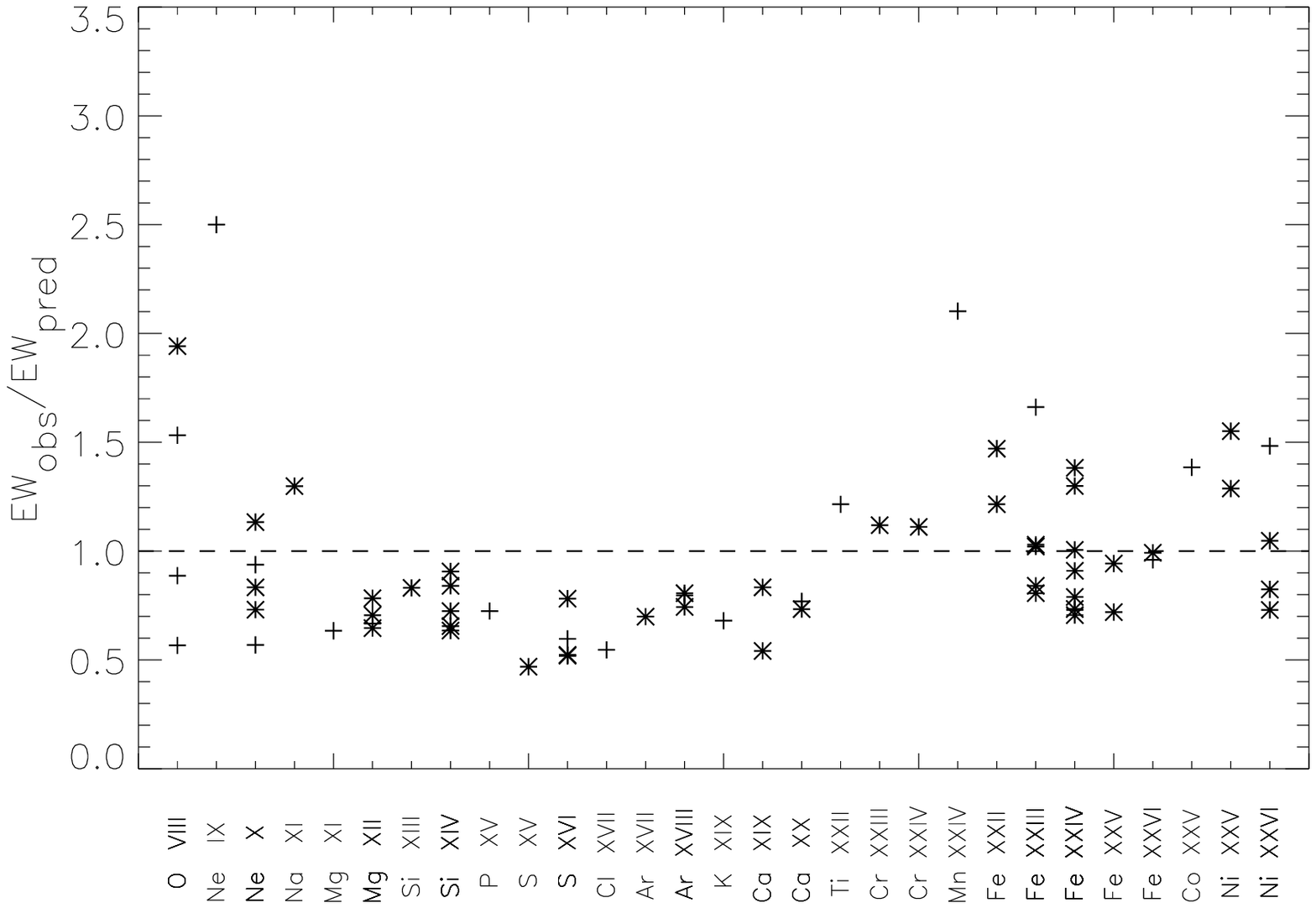,width=12.0cm}}
\end{figure*}
\begin{figure*}
\end{figure*}
\clearpage

\normalsize

\begin{thebibliography}{1}

\bibitem{ss} Shakura, N. I., Sunyaev, R. A., Balck holes in binary
systems.  Observational Appearance.  {\it Astron. Astrophys.}, {\bf
24}, 337-355 (1973).

\bibitem{bh91} Balbus, S. A., Hawley, J. F., A powerful local shear
instability in weakly magnetized disks.  {\it Astrophys. J.}, {\bf 376},
214-233 (1991).

\bibitem{hgb95} Hawley, J. F., Gammie, C. F., Balbus, S. A., Local
Three-dimensional Magnetohydrodynamic Solutions of Accretion
Disks. {\it Astrophys. J.}, {\bf 440}, 742-763 (1995).

\bibitem{bh98} Balbus, S. A., Hawley, J. F., Instability, turbulence,
and enhanced transport in accretion disks.  {\it Rev. Mod. Phys.},
{\bf 70}, 1-53 (1998).

\bibitem{bp} Blandford, R. D.,Payne, D. G, Hydromagnetic flows from
accretion disks and the production of radio
jets. {\it Mon. Not. R. Astron. Soc.}, {\bf 199}, 883-903 (1982).

\bibitem{orosz} Orosz, J., Bailyn, C. D., Optical Observations of GRO
J1655$-$40 in Quiescence. I. A Precise Mass for the Black Hole
Primary.  {\it Astrophys. J.}, {\bf 477}, 876-896 (1997).

\bibitem{hr} Hjellming, R. M., Rupen, M. P., episodic ejection of
relativistic jets by the X-ray transient GRO J1655$-$40, {\it Nature},
{\bf 375}, 464-468 (1995).

\bibitem{m04} Miller, J. M., et al., Chandra/HETGS Spectroscopy of the
Galactic Black Hole GX 399$-$4: A Relativistic Iron Emission Line and
Evidence for a Seyfert-like Warm Absorber.  Astrophys. J., {\bf
601}, 450-465 (2004).

\bibitem{houck} Houck, J. C., Denicola, L. A., ISIS: An Interactive
Spectral interpretation System for High Resolution X-ray Spectroscopy.
{\it Astronomical Data Analysis Softweare and Systems IX, Astronomical
Society of the Pacific Conference Proceedings}, {\bf 216}, 591-594
(2000).

\bibitem{dl} Dickey, J. M., Lockman, F. J., H I in the Galaxy.  {\it
Annu. Rev. Astron. Astrophys.}, {\bf 28}, 215-261 (1990).

\bibitem{vvf} Verner, D. A., Verner, E. M., Ferland, G. J., Atomic
Data for Permitted Resonance Lines of Atoms and Ions from H to Si, and
S, Ar, Ca, and Fe.  {\it Atomic Data and Nuclear Data Tables}, {\bf
64}, 1 (1996).

\bibitem{nist} The National Institute of of Standards and Technology
(NIST) Atomic Spectra Database, Standard Reference Database 78,
available on-line at http:\/\/physics.nist.gov\/cgi-bin\/AtData\/main\_asd (2005).

\bibitem{nah} Nahar, S., Pradhan, A. K., Atomic data from the Iron
Project.  XXXV.  Relativistic fine structure oscillator strengths for
Fe XXIV and Fe XXV.  {\it Astron. Astrophs. Suppl.}, {\bf 135},
347-357 (1999).

\bibitem{jr93} Raymond, J., A model of an X-ray-illuminated accretion
disk and corona.  {\it Astrophys. J.}, {\bf 412}, 267-277 (1993).

\bibitem{colgan} Colgan, J., Pindzola, M. S., Badnell, N. R.,
Dielectronic recombination data for dynamic finite-density plasmas. V:
the lithium isoelectronic sequence.  {\it Astron. Astrophys}, {\bf
417}, 1183-1188 (2004).  

\bibitem{sp78} Spitzer, L., {\it Physical Processes in the
Interstellar Medium}.  New York: Wiley (1978).

\bibitem{grev} Grevesse, N., \& Sauval, A. J., in {\it Solar
Composition and its Evolution -- from Core to Corona},
ed. C. Fr\"{o}lich, M. C. E. Huber, S. K. Solanski, \& r. von Steiger,
(Dordrecht: Kluwer), 161 (1998).

\bibitem{vrt} Vrtilek, S., et al., Observations of Cygnus X-2 with IUE
- Ultraviolet results from a multiwavelength campaign.  {\it
Astron. Astrophys.}, {\bf 234}, 162-173 (1990).

\bibitem{beg83} Begelman, M. C., McKee, Shields, G. S., Compton heated
winds and coronae above accretion disks.  II Dynamics.  {\it
Astrophys. J.}, {\bf 271}, 70-89 (1983).

\bibitem{cn05} Chelouche, D., Netzer, H., Dynamical and Spectral
Modeling of the Ionized Gas and Nuclear Environment in NGC 3783. {\it
Astrophys. J.}, {\bf 625}, 95-107 (2005).

\bibitem{psk00} Proga, D., Stone, J. M., Kallman, T. R., Dynamics of
Line-Driven winds in Active Galactic Nuclei.  {\it Astrophys. J.},
{\bf 543}, 686-696 (2000).

\bibitem{ms00} Miller, K. A., Stone, J. M., The Formation and
Structure of a Strongly Magnetized Corona above a Weakly Magnetized
Accretion Disk.  {\it Astrophys. J.}, {\bf 534}, 398-419 (2000).

\bibitem{spruit} Spruit, H. C., in ``Physical Processes in
Binary Stars'', eds. R. A. M. J. Wijers, M. B. Davies, and C. A. Tout,
Kluwer Dordrecht (NATO ASI series) (1996).

\bibitem{proga03} Proga, D., Numerical Simulations of Mass Outflows
Driven from Accretion Disks by Radiation and Magnetic Forces.  {\it
Astrophys. J.}, {\bf 585}, 406-417 (2003).

\bibitem{calvet} Calvet, N., Hartmann, L., Kenyon, S. J., Mass loss
from pre-main-sequence accretion disks. I - The accelerating wind of
FU Orionis.  {\it Astrophys. J.}, {\bf 402}, 623-634 (1993).

\bibitem{sb02} Schulz, N. S., Brandt, W. N., Variability of the X-ray
P Cygni Line Profiles from Cicinus X-1 Near Zero Phase''.  {\it
Astrophys. J.}, {\bf 572}, 972-983 (2002).

\bibitem{mauche} Mauche, C. W., \& Raymond, J. C., Extreme Ultraviolet
Explorer Observations of OY Carinae in Superoutburst.  {\it
Astrophys. J.}, {\bf 541}, 924-936 (2000).

\bibitem{kk94} Konigl, A., Kartje, J. F., Disk-Driven Hydromagnetic
Winds as a Key Ingredient of Active Galactic Nuclei Unification
Schemes.  {\it Astrophys. J.}, {\bf 434}, 446-467 (1994).

\bibitem{e05} Everett, J. E., Radiative Transfer and Acceleration in
Magnetocentrifugal Winds.  {\it Astrophys. J.}, {\bf 631}, 689-706
(2005).

\end{thebibliography}
\end{document}